\documentclass{PoS} 

\title{Critical behavior and continuum scaling of $3D$ $Z(N)$ lattice gauge 
theories}

\ShortTitle{Critical behavior of $3D$ $Z(N)$ LGTs}

\author{Oleg Borisenko\\
        BITP, National Academy of Sciences of Ukraine,
        03680 Kiev, Ukraine \\
        E-mail: \email{oleg@bitp.kiev.ua}}

\author{Volodymyr Chelnokov\\
        BITP, National Academy of Sciences of Ukraine,
        03680 Kiev, Ukraine\\
        E-mail: \email{chelnokov@bitp.kiev.ua}}

\author{Mario Gravina\\
        Dipartimento di Fisica, Universit\`a della Calabria,\\
        \& INFN - Gruppo Collegato di Cosenza, I-87036 Rende, Italy\\
        E-mail: \email{gravina@cs.infn.it}}

\author{\speaker{Alessandro Papa}\\
        Dipartimento di Fisica, Universit\`a della Calabria,\\
        \& INFN - Gruppo Collegato di Cosenza, I-87036 Rende, Italy\\
        E-mail: \email{papa@cs.infn.it}}

\abstract{Three-dimensional $Z(N)$ lattice gauge theories are studied 
numerically at finite temperature for $N$ = 5, 6, 8, 12, 13, 20 and for 
$N_t$=2,4,8. 
For each model the location of phase transitions and its critical
indices are determined. The scaling of critical points with $N$ is proposed.
The data obtained enable us to verify the scaling near the continuum limit
for the $Z(N)$ models at finite temperatures.}

\FullConference{The 32nd International Symposium on Lattice Field Theory\\
		 23-28 June, 2014\\
		 Columbia University New York, NY}

\begin{document}

\section{Introduction} 

In this paper we focus on the phase structure of $Z(N)$ lattice gauge theories
(LGTs), which are interesting on their own and can provide for useful insights 
into the universal properties of $SU(N)$ LGTs, being $Z(N)$ the center subgroup 
of $SU(N)$. The most general action for the $Z(N)$ LGT is
\begin{equation} 
S_{\rm gauge} \ = \ \sum_x \sum_{n<m} \ \sum_{k=1}^N \beta_k
\cos \left( \frac{2 \pi k}{N} \left(s_n(x) + s_m(x+e_n) 
-s_n(x+e_m) - s_m(x) \right) \right) \ .
\label{action_gauge}
\end{equation}
Gauge fields take on values $s_n(x)=0,1,\cdots,N-1$ and are defined on the links
of the lattice. $Z(N)$ gauge models can generally be divided into two classes: 
{\em standard Potts} models (all $\beta_k$ equal) and {\em vector} models 
(otherwise). The conventional vector model corresponds to $\beta_k=0$ for all 
$k>1$. For $N=2,\ 3$ the Potts and vector models are equivalent. 

An extended description of the phase structure of $Z(N)$ LGTs in three 
dimension can be found in~\cite{3d_zn_strcoupl,ZN_fin_T,ZN_zero_T}.
In those papers we explored the phase structure of the vector 
$Z(N)$ LGTs for $N>4$. We considered first an anisotropic lattice in the 
limit where the spatial coupling vanishes~\cite{3d_zn_strcoupl} and were
able to present both renormalization group (RG) and numerical evidences
for the existence of two BKT-like phase transitions: a 
(i) {\em first transition}, from a symmetric, confining phase to
an intermediate phase, where the $Z(N)$ symmetry is enhanced to $U(1)$ 
symmetry; (ii)  a {\em second transition}, from the intermediate phase to a 
phase with broken $Z(N)$ symmetry. We computed also some critical indices, 
which appear to agree with the corresponding indices of $2D$ 
$Z(N)$ spin models, thus giving further support to the Svetitsky-Yaffe 
conjecture~\cite{Svetitsky}. In particular, we found that the magnetic
critical index $\eta$ at the first transition, $\eta^{(1)}$, takes the
value 1/4 as in $2D$ $XY$, while its value at the second transition, 
$\eta^{(2)}$, is equal to $4/N^2$. Then, we extended our analysis to the full 
isotropic $3D$ $Z(N)$ LGT at finite temperature~\cite{ZN_fin_T} and
confirmed by numerical Monte Carlo simulations~\cite{ZN_fin_T} that the full 
gauge models with $N>4$ possess two phase transitions of the BKT type, with 
critical indices coinciding with those of $2D$ vector spin models. 

Here we extend the study of Ref.~\cite{ZN_fin_T} to other values of $N$ and 
to $N_t=8$ and aim at checking the scaling near the continuum limit and
at establishing the scaling formula for critical points with $N$. 
In particular, the theory of dimensional cross-over~\cite{caselle} explains 
how critical couplings and indices of a finite temperature LGT (finite $N_t$) 
approach critical couplings and indices of the corresponding zero-temperature 
theory ($N_t \to \infty$). This provides us with a way to crosscheck our 
zero-temperature results~\cite{ZN_zero_T} and thus predict the critical 
temperature in the continuum limit.

The standard approach for studying a BKT transition consists in using Binder 
cumulants and susceptibilities of the Polyakov loop to determine critical 
couplings and critical indices. Here, as in Ref.~\cite{ZN_fin_T}, we follow a 
different strategy: we move to a dual formulation and use Binder cumulants and 
susceptibilities of {\em dual $Z(N)$ spins}. This implies that (i) the critical 
behavior of dual spins is reversed with respect to that of Polyakov loops, 
namely the spontaneously-broken ordered phase is mapped to the symmetric phase 
and {\it vice versa}; (ii) the magnetic critical indices $\eta$ are 
interchanged, whereas the index $\nu$ is expected to be the same (=1/2) at both 
transitions (see Ref.~\cite{ZN_fin_T} for details). The obvious advantage of 
this approach is that cluster algorithms become available, with considerable 
speed up in the numerical procedure. 

\section{Theoretical setup}

The $3D$ $Z(N)$ gauge theory on an anisotropic $3D$ lattice $\Lambda$ can 
generally be defined as 
\begin{equation}
Z(\Lambda ;\beta_t,\beta_s;N) \ = \  \prod_{l\in \Lambda}
\left ( \frac{1}{N} \sum_{s(l)=0}^{N-1} \right ) \ \prod_{p_s} Q(s(p_s)) \
\prod_{p_t} Q(s(p_t)) \; ,
\label{PTdef}
\end{equation}
where the link angles $s(l)$ are combined into the conventional plaquette angle
\begin{equation}
s(p) \ = \ s_n(x) + s_m(x+e_n) - s_n(x+e_m) - s_m(x) \ .
\label{plaqangle}
\end{equation}
Here, $e_n$  ($n=0,1,2$) denotes a unit vector in the $n$-th direction and
the notation $p_t$ ($p_s$) stands for the temporal (spatial) plaquettes. 
Periodic boundary conditions (BC) on gauge fields are imposed in all 
directions. The most general $Z(N)$-invariant Boltzmann weight with $N-1$ 
different couplings is
\begin{equation}
Q(s) \ = \
\exp \left [ \sum_{k=1}^{N-1} \beta_p(k) \cos\frac{2\pi k}{N}s \right ] \ .
\label{Qpgen}
\end{equation}
The Wilson action corresponds to the choice $\beta_p(1)=\beta_p$, 
$\beta_p(k)=0, k=2,...,N-1$, which is the one adopted in this work. 
Furthermore, we will consider an isotropic lattice: $\beta_s=\beta_t=\beta$.

Our study is based on the mapping of the gauge model to a generalized $3D$ 
$Z(N)$ spin model on a dual lattice $\Lambda_d$, whose action is
\begin{equation}
\label{modaction}
S \ =\ \sum_{x}\ \sum_{n=1}^3 \sum_{k = 1}^{N-1} \ \beta_k \  
\cos \left( \frac{2 \pi k}{N} \left(s(x) - s(x+e_n) \right) \right) \ .
\end{equation}
The dual mapping is realized once one specifies the relationship between the 
original gauge coupling $\beta$ and the dual effective couplings $\beta_k$.
This has been done in Ref.~\cite{ZN_fin_T} (see also Ref.~\cite{ukawa}) and
the result is
\begin{equation}
\beta_k \ =\ \frac{1}{N} \sum_{p = 0}^{N - 1} \ln \left [ \frac{Q_d(p)}
{Q_d(0)} \right ] \  \cos \left(\frac{2 \pi p k}{N} \right) \ .
\label{couplings}
\end{equation}

For $N=5$ it can be seen explicitly~\cite{ZN_fin_T} that $|\beta_1|\gg 
|\beta_2|$, thus suggesting that the $3D$ vector spin model with only $\beta_1$
non-vanishing gives already a reasonable approximation of the gauge model.
Moreover the weak and the strong coupling regimes are interchanged, {\it i.e.} 
when $\beta\to\infty$ the effective couplings $\beta_k\to 0$ and, therefore, 
the ordered symmetry-broken phase is mapped to a symmetric phase with vanishing 
magnetization of dual spins, whereas the symmetric phase at small $\beta$ 
becomes an ordered phase where the dual magnetization is non-zero. The
interchange of phases under the dual mapping is not a special feature of $N=5$, 
but is rather a general property valid for any $N$. 
In Ref.~\cite{ZN_fin_T} it was also discussed that at the
critical point $\beta_{\rm c}^{(1)}$ of the first transition of the LGT 
(from the symmetric to the intermediate phase), the {\em dual} 
correlation function scales with a critical index $\eta$ equal to the
index $\eta^{(2)}=4/N^2$ of the Polyakov loop correlator in the LGT, while
at the critical point $\beta_{\rm c}^{(2)}$ of the second transition
in the LGT (from the intermediate to the broken phase), it scales with 
a critical index $\eta$ equal to the index $\eta^{(1)}=1/4$ of the Polyakov 
loop correlator in the LGT. This can be proved in the Villain formulation of
the $2D$ theory and only conjectured (but confirmed numerically) in the $3D$ 
case~\cite{ZN_fin_T}.

\section{Numerical setup and results}

The $3D$ $Z(N)$ spin model, dual of the $3D$ $Z(N)$ Wilson LGT, has been
simulated by means of a cluster algorithm on $N_t \times L \times L$ lattices 
with periodic BC. The system has been studied for $N$ = 5, 6, 8, 12, 13
and 20 on lattices with the temporal extension $N_t$=2, 4, 8. With respect 
to our previous work~\cite{ZN_fin_T}, we considered new values of 
$N$ (6, 8, 12, 20) and included also $N_t=8$. We focused on the following 
observables:
\begin{itemize}
\item complex magnetization $M_L = |M_L| e^{i \psi}$, with
$M_L \ =\  \sum_{x \in \Lambda} \exp \left( \frac{2 \pi i}{N} s(x) \right)$,
where we stress that $s(x)$ is a dual spin variable;
\item real part of the rotated magnetization, $M_R = |M_L| \cos(N \psi)$,
and normalized rotated magnetization, $m_\psi = \cos(N \psi)$;
\item susceptibilities of $M_L$ and $M_R$, $\chi_L^{(M)}$, $\chi_L^{(M_R)}$:
\ \ $\chi_L^{(\mathbf\cdot)} \ =\  L^2 N_t \left(\left< \mathbf\cdot^2 \right> 
- \left< \mathbf\cdot \right>^2 \right)$;
\item Binder cumulants $U_L^{(M)}$ and $B_4^{(M_R)}$:
\ \ $U_L^{(M)}\ =\ 1 - \frac{\left\langle \left| M_L \right| ^ 4 
\right\rangle}{3 \left\langle \left| M_L \right| ^ 2 \right\rangle^2}\;, 
\;\;\;\;\;
B_4^{(M_R)}\ =\ \frac{\left\langle \left| M_R 
- \left\langle M_R \right\rangle \right| ^ 4 \right\rangle}
{\left\langle \left| M_R - \left\langle M_R \right\rangle \right| ^ 2 
\right\rangle ^ 2 }$.
\end{itemize}

To determine the critical couplings of the {\em second transition point}, 
$\beta_{\rm c}^{(2)}$, we have looked for the value of $\beta$ at which the curves
giving the Binder cumulant $U_L^{(M)}(\beta)$ on lattices with different 
size $L$ ``intersect'' (see Ref.~\cite{ZN_fin_T2} for details).
The same method can in principle be used for the couplings of the
{\em first transition}, $\beta_{\rm c}^{(1)}$, using either the Binder cumulant 
$B_4^{(M_R)}$ or $m_\psi$; it turned out, however, that the precision required 
by this method on these observables could not be met with a sensible simulation 
time. For this reason, as the position of the first critical point we used our 
previous determinations given in Ref.~\cite{ZN_fin_T}, where 
$\beta_{\rm c}^{(1)}$ was taken as the value of $\beta$ at which 
$B_4^{(M_R)}$ and $m_\psi$ plotted {\it versus} $(\beta-\beta_{\rm c}^{(1)}) 
{\ln L}^{1/\nu}$ show the best overlap for different values of $L$. 
The results of the determinations of $\beta_{\rm c}^{(1)}$ and 
$\beta_{\rm c}^{(2)}$ are summarized in Table~\ref{tbl:crit_betas}.

\begin{table}[tb]
\caption[]{Values of $\beta_{\rm c}^{(1)}$ and $\beta_{\rm c}^{(2)}$ obtained 
for various $N_t$ in $3D$ $Z(N)$ with $N = 5,\ 6,\ 8,\ 12,\ 13$ and 20.}
%\scriptsize
\small
\begin{center}
\begin{tabular}{|c|c|c|c|}
\hline
$N$ & $N_t$ & $\beta_{\rm c}^{(1)}$ & $\beta_{\rm c}^{(2)}$ \\
\hline
 5 & 2 & 1.617(2) & 1.6972(14) \\ 
 5 & 4 & 1.943(2) & 1.9885(15) \\ 
 5 & 6 & 2.05(1)  & 2.08(1) \\
 5 & 8 & 2.085(2) & 2.1207(9) \\ 
 5 & 12 & 2.14(1) & 2.16(1) \\
\hline
 6 & 2 & -        & 2.3410(15) \\ 
 6 & 4 & -        & 2.725(12) \\ 
 6 & 8 & -        & 2.899(4) \\ 
\hline
 8 & 2 & -        & 3.8640(10)\\ 
 8 & 4 & 2.544(8) & 4.6864(15) \\ 
 8 & 8 & 3.422(9) & 4.9808(5) \\ 
\hline
\end{tabular}
\hspace{2cm}
\begin{tabular}{|c|c|c|c|}
\hline
$N$ & $N_t$ & $\beta_{\rm c}^{(1)}$ & $\beta_{\rm c}^{(2)}$ \\
\hline
12 & 2 & -        & 8.3745(5) \\ 
12 & 4 & -        & 10.240(7) \\ 
12 & 8 & -        & 10.898(5) \\ 
\hline
 13 & 2 & 1.795(4) & 9.735(4) \\ 
 13 & 4 & 2.74(5)  & 11.959(6) \\
 13 & 8 & 3.358(7) & 12.730(2) \\
\hline
 20 & 2 & -       & 22.87(4)   \\
 20 & 4 & 2.57(1) & 28.089(3) \\ 
 20 & 8 & 3.42(5) & 29.758(6)   \\ 
\hline
\end{tabular}
\end{center}
\label{tbl:crit_betas}
\end{table}

\begin{figure}[tb]
\centering
\includegraphics[width=0.32\textwidth]{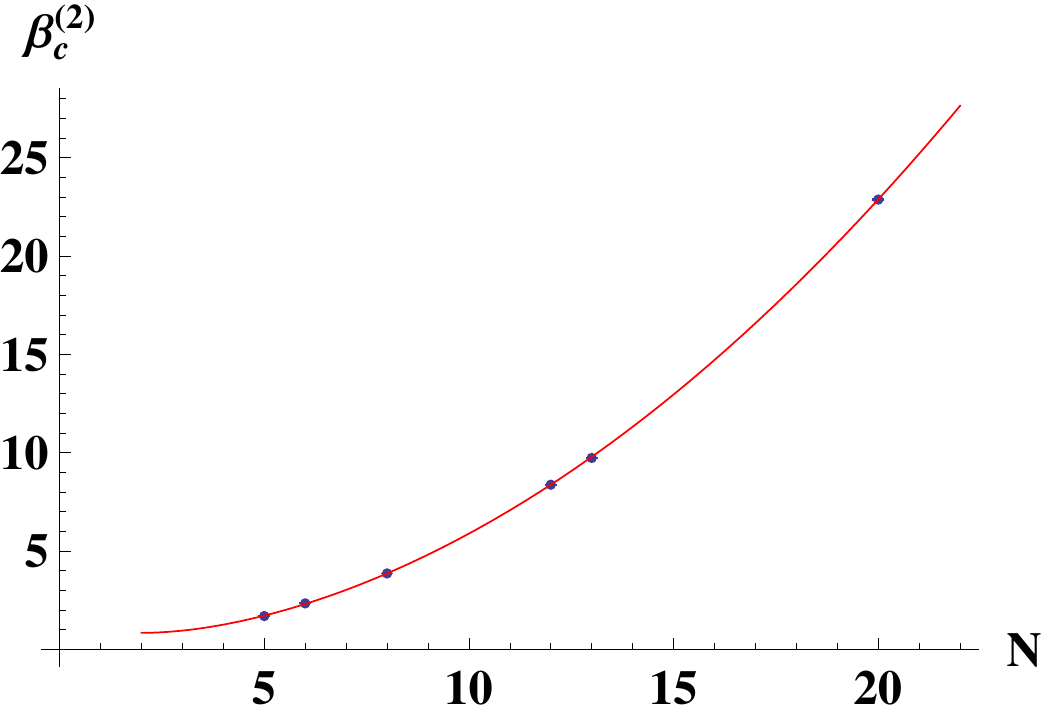}
\includegraphics[width=0.32\textwidth]{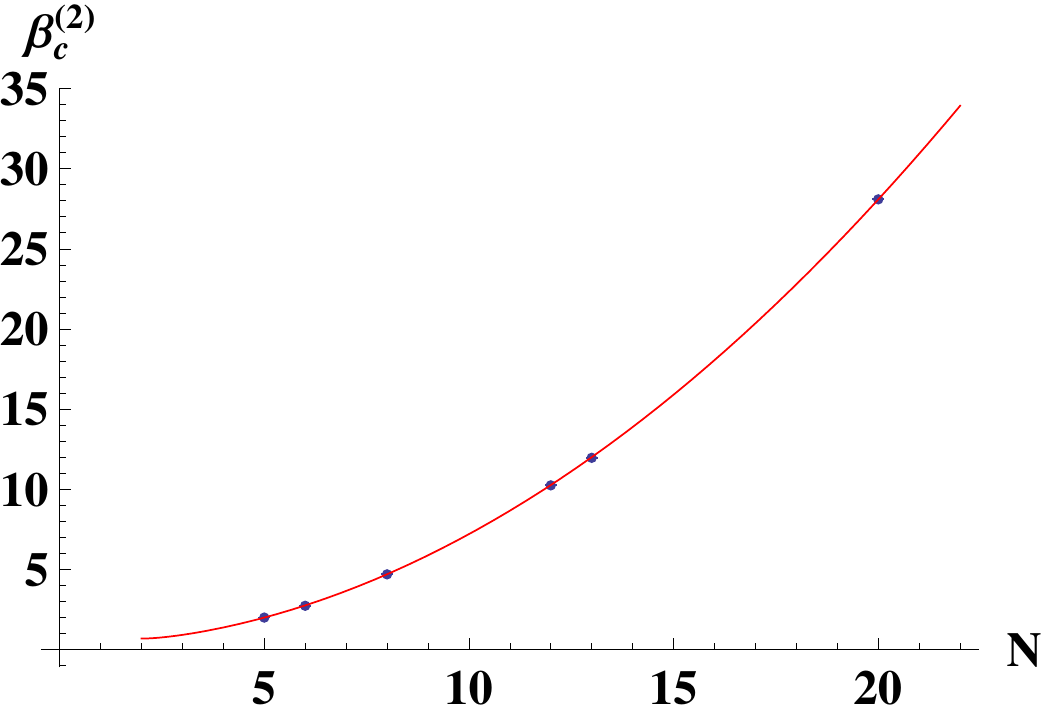}
\includegraphics[width=0.32\textwidth]{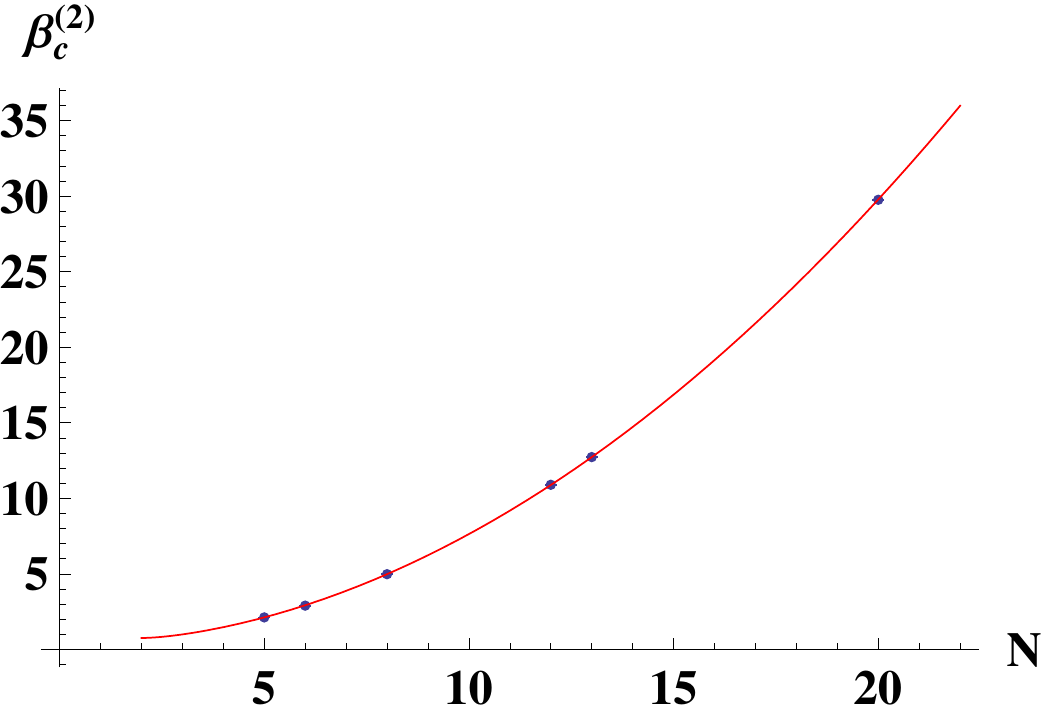}
\caption[]{Scaling function $A/(1-\cos{2\pi/N}) + B (1-\cos{2\pi/N})$
{\it versus} data for $\beta_{\rm c}^{(2)}$ at $N_t=2,\ 4,\ 8$ (from left to 
right).}
\label{fig:ndep2}
\end{figure}

For the critical couplings at the second transition, $\beta_{\rm c}^{(2)}$,
where determinations for many values of $N$ are available, we tried to
find a simple scaling dependence with $N$ at fixed $N_t$. 
From the solution of the renormalization group equations for $2D$
$Z(N)$ spin model, we know that in that model $\beta_{\rm c}^{(2)}(N)$ grows 
as $N^2$ for large $N$~\cite{2d_zn}. In~\cite{3d_zn_strcoupl} we have found 
that this is the case also for the $3D$ $Z(N)$ LGT at finite temperature, at 
least in the strong coupling limit. Taking inspiration from Ref.~\cite{bhanot}, 
we started from a scaling law written in the form 
$\beta_{\rm c}^{(2)}(N) = A/(1-\cos{2\pi/N})$. Then, considering that 
the next non-negligible correction comes at the order $1/N^2$, we added a 
second term and ended up with the same scaling function we used in 
the zero-temperature case~\cite{ZN_zero_T},
\[
\beta_{\rm c}^{(2)}(N) = \frac{A}{(1-\cos{2\pi/N})} + B (1-\cos{2\pi/N})\;.
\]
In Table~\ref{tbl:ndep2} we report the values of the parameters $A$ and $B$
for $N_t=2,\ 4,\ 8$, while Figs.~\ref{fig:ndep2} shows the fitting functions 
against numerical data. 

\begin{table}[tb]
\caption{Parameters of the scaling with $N$ of the second transition point, 
$\beta_{\rm c}^{(2)} = A /{(1-\cos{2\pi/N})} + B (1-\cos{2\pi/N})$ at fixed 
$N_t$.}
\label{tbl:ndep2}
%\scriptsize
\small
\begin{center}
\begin{tabular}{||c||c|c|c||}
\hline
$N_t$ & $A$      & $B$         & $\chi^2_{\rm r}$ \\
\hline                                        
 2 & 1.1194(11)  & 0.141(24)   & 209  \\
 4 & 1.37440(60) & -0.0046(88) & 18.2 \\
 8 & 1.45745(57) & 0.0155(53)  & 16.1 \\
\hline
\end{tabular}
\end{center}
\end{table}

Finding the continuum limit of the finite temperature theory in the first
or in the second transition amounts to extrapolate the corresponding critical
couplings, $\beta_{\rm c}^{(1)}$ or $\beta_{\rm c}^{(2)}$, to the limit
$N_t\to\infty$ at fixed $N$.
The theory of dimensional cross-over~\cite{caselle} suggests the fitting
function to be used:
\begin{equation}
\beta_{\rm c}^{(1,2)}(N_t) = \beta_{{\rm c},\ T=0}^{(1,2)}
- (N_t a T_{\rm c})^{-1/\nu}\;,
\label{fit_cont}
\end{equation}
where $\beta_{{\rm c},\ T=0}^{(1,2)}$ and $\nu$ are the critical couplings 
and the critical index of the zero-temperature theory. Since we know that,
for any $N$, the $3D$ $Z(N)$ LGT at zero-temperature exhibits only one
phase transition, with the critical index $\nu$ depending on the side from
which the transition is approached~\cite{ZN_zero_T}, we expect that, for
a given $N$, the fit parameters $\beta_{{\rm c},\ T=0}^{(1)}$ and 
$\beta_{{\rm c},\ T=0}^{(2)}$ take the same value and agree with the 
zero-temperature critical coupling at the same $N$. 
As for the fit parameter $\nu$, we expect it to agree with the value of the 
critical index $\nu$ at one of the two sides of the zero-temperature transition.
We fitted with the function given in~(\ref{fit_cont}) our data for the
critical couplings $\beta_{\rm c}^{(1)}(N_t)$ at $N$=5 and for the critical 
couplings $\beta_{\rm c}^{(2)}(N_t)$ at $N$=5, 6, 8, 12, 13, 20 (see 
Table~\ref{tbl:clim}). In some cases 
in the fit we fixed either $\beta_{{\rm c},\ T=0}^{(1,2)}$ or $\nu$, or both, at the
values known from the zero-temperature theory~\cite{ZN_zero_T}. The scenario 
which emerges from the inspection of Tables~\ref{tbl:clim} is that, despite 
the large reduced chi-squared obtained
in a few cases, the agreement between the fit parameters 
$\beta_{{\rm c},\ T=0}^{(1,2)}$ and the known zero-temperature critical 
couplings~\cite{ZN_zero_T} is satisfactory. As for the value of the fit
parameter $\nu$, results are not precise enough to discriminate
between the known values of the critical index $\nu$ of the zero-temperature 
theory at one or the other side of the transition~\cite{ZN_zero_T}.
This analysis allows us for the determination of the critical temperature 
$a T_{\rm c}$ in the continuum limit for all the values of $N$ considered
in this work.

\begin{table}[tb]
\caption[]{Results of the fit of $\beta_{\rm c}^{(1)}(N_t)$ for $N=$ and of 
$\beta_{\rm c}^{(2)}(N_t)$ for $N$=5, 6, 8, 12, 13, 20 with the 
function~(\ref{fit_cont}). Parameters are given without errors when their 
values were fixed at the known results of the $T=0$ corresponding 
theory~\cite{ZN_zero_T} (for the $\nu$ index we considered both the values at 
the left and at the right of the $T=0$ critical point). Parameters are given 
with a (-) mark when their errors are unavailable and with a ${}^{*}$ mark when 
obtained from fits on data with $N_t=4,\ 8$ only (in general, $N_t=2,\ 4,\ 8$ 
were considered).}
\label{tbl:clim}
\footnotesize
\begin{center}
\begin{tabular}{|c|c|c|c|c|}
\hline
$N$ & $aT_{\rm c}$ & $\beta_{{\rm c},\ T=0}^{(1)}$ & $\nu$ & $\chi_{\rm r}^2$\\
\hline
    & 0.790(5)  & 2.198(9)  & 0.84(3)   & 1.21  \\
    & 0.764(14) & 2.144(9)  & 0.670     & 23.1  \\
    & 0.758(16) & 2.135(11) & 0.640     & 33.6  \\ 
5   & 0.786(7)  & 2.17961   & 0.788(10) & 2.66  \\
    & 0.722(16) & 2.17961   & 0.670     & 105   \\
    & 0.709(19) & 2.17961   & 0.640     & 171   \\ 
\hline     
\end{tabular}

\vspace{0.2cm}

\begin{tabular}{|c|c|c|c|c|}
\hline
$N$ & $aT_{\rm c}$ & $\beta_{{\rm c},\ T=0}^{(2)}$ & $\nu$ & $\chi_{\rm r}^2$\\
\hline
    & 0.868(-)          & 2.23055(-) & 0.877(-)  & -            \\
    & 0.813(27)         & 2.177(12)  & 0.670     & 158          \\
    & 0.803(30)         & 2.170(14)  & 0.640     & 223          \\ 
5   & 0.825(38)         & 2.17961    & 0.692(45) & 131          \\
    & 0.810(13)         & 2.17961    & 0.670     & 81.8         \\
    & 0.776(31)${}^{*}$ & 2.17961    & 0.670     & 74.2${}^{*}$ \\
    & 0.789(17)         & 2.17961    & 0.640     & 161          \\ 
    & 0.731(18)${}^{*}$ & 2.17961    & 0.640     & 31.4${}^{*}$ \\ 
\hline                      
    & 0.6814(-)         & 3.04317(-) & 0.876(-)  & -            \\
    & 0.6769(76)        & 2.977(10)  & 0.674     & 5.02         \\
    & 0.6740(85)        & 2.969(12)  & 0.642     & 6.90         \\ 
6   & 0.6832(46)        & 3.00683    & 0.768(15) & 1.14         \\
    & 0.6573(47)        & 3.00683    & 0.674     & 22.6         \\
    & 0.572(13)${}^{*}$ & 3.00683    & 0.674     & 1.44${}^{*}$ \\
    & 0.6487(60)        & 3.00683    & 0.642     & 40.6         \\ 
    & 0.542(21)${}^{*}$ & 3.00683    & 0.642     & 4.48${}^{*}$ \\ 
\hline
    & 0.42330(-)        & 5.14422(-) & 0.674(-)  & -            \\
    & 0.42378(12)       & 5.14299(25)& 0.672     & 0.19         \\
    & 0.4316(22)        & 5.1225(46) & 0.637     & 66.5         \\ 
8   & 0.4294(12)        & 5.12829    & 0.648(6)  & 33.0         \\
    & 0.4287(39)        & 5.12829    & 0.672     & 321          \\
    & 0.4427(39)${}^{*}$& 5.12829    & 0.672     & 177${}^{*}$  \\
    & 0.4298(19)        & 5.12829    & 0.637     & 86.1         \\ 
    & 0.4216(10)${}^{*}$& 5.12829    & 0.637     & 2.21${}^{*}$ \\ 
\hline     
\end{tabular}
%\hspace{1cm}
\begin{tabular}{|c|c|c|c|c|}
\hline
$N$ &$aT_{\rm c}$ & $\beta_{{\rm c},\ T=0}^{(2)}$ & $\nu$ & $\chi_{\rm r}^2$ \\
\hline
    & 0.24728(-)         & 11.2566(-)  & 0.674(-)  & -            \\
    & 0.24559(13)        & 11.2640(23) & 0.670     & 0.22         \\
    & 0.25615(72)        & 11.218(12)  & 0.640     & 6.18         \\ 
12  & 0.2602(32)         & 11.1962     & 0.630(11) & 14.2         \\
    & 0.24954(28)        & 11.1962     & 0.670     & 89.8         \\
    & 0.2619(87)${}^{*}$ & 11.1962     & 0.670     & 55.5${}^{*}$ \\
    & 0.25742(10)        & 11.1962     & 0.640     & 12.7         \\ 
    & 0.2597(51)${}^{*}$ & 11.1962     & 0.640     & 21.3${}^{*}$ \\ 
\hline
    & 0.22433(-)         & 13.1391(-)  & 0.654(-)  & -            \\
    & 0.21872(53)        & 13.1656(56) & 0.671     & 5.88         \\
    & 0.22851(40)        & 13.1199(42) & 0.642     & 3.40         \\ 
13  & 0.2310(12)         & 13.1077     & 0.635(4)  & 8.86         \\
    & 0.2225(30)         & 13.1077     & 0.671     & 314          \\
    & 0.2342(62)${}^{*}$ & 13.1077     & 0.671     & 113${}^{*}$  \\
    & 0.22928(67)        & 13.1077     & 0.642     & 16.0         \\ 
    & 0.2311(24)${}^{*}$ & 13.1077     & 0.642     & 19.2${}^{*}$ \\ 
\hline
    & 0.144857(-)        & 30.5427(-)  & 0.608(-)  & -            \\
    & 0.1297(37)         & 30.73(10)   & 0.673     & 147          \\
    & 0.1356(24)         & 30.658(64)  & 0.647     & 58.8         \\ 
20  & 0.1357(26)         & 30.6729     & 0.642(19) & 58.2         \\
    & 0.13171(98)        & 30.6729     & 0.673     & 97.3         \\
    & 0.13199(13)${}^{*}$& 30.6729     & 0.673     & 1.57${}^{*}$ \\
    & 0.13506(54)        & 30.6729     & 0.647     & 31.0         \\ 
    & 0.13519(49)${}^{*}$& 30.6729     & 0.647     & 23.9${}^{*}$ \\ 
\hline
\end{tabular}
\end{center}
\end{table}   
              
%\begin{table}[tb]
%\caption[]{Same as Table~\ref{tbl:clim2} for $\beta_{\rm c}^{(1)}(N_t)$ for 
%$N$=5.}
%\label{tbl:clim1}
%\scriptsize
%\begin{center}
%\begin{tabular}{|c|c|c|c|c|}
%\hline
%$N$ & $aT_{\rm c}$ & $\beta_{{\rm c},\ T=0}^{(1)}$ & $\nu$ & $\chi_{\rm r}^2$\\
%\hline
%    & 0.790(5)  & 2.198(9)  & 0.84(3)   & 1.21  \\
%    & 0.764(14) & 2.144(9)  & 0.670     & 23.1  \\
%    & 0.758(16) & 2.135(11) & 0.640     & 33.6  \\ 
%5   & 0.786(7)  & 2.17961   & 0.788(10) & 2.66  \\
%    & 0.722(16) & 2.17961   & 0.670     & 105   \\
%    & 0.709(19) & 2.17961   & 0.640     & 171   \\ 
%\hline     
%\end{tabular}
%\end{center}
%\end{table}   

Some critical indices at the two transitions in the $3D$ $Z(N)$ LGT at
finite temperature can be extracted by the standard FSS analysis.
In particular, the behavior on the lattice size $L$ of the standard
magnetization $M_L$ and of its susceptibility at the second transition
allows to extract the indices $\beta/\nu$ and $\gamma/\nu$ through a
fit with the functions
\begin{equation}
M_L = A_{M_L} L^{-\beta/\nu} \;, \;\;\;\;\;
\chi_{M_L} = A_{\chi_{M_L}} L^{\gamma/\nu} \;.
\end{equation}
Similarly, the behavior on $L$ of the rotated magnetization $M_R$ and of
its susceptibility at the first transition point allow the extraction 
of the same critical indices at that transition.
Thereafter, the hyperscaling relation $2 \beta/\nu + \gamma/\nu = 2$ can be
checked and the magnetic index $\eta= 2 - \gamma/\nu$ can be extracted
at both transitions.
Our results are reported in Ref.~\cite{ZN_fin_T2} and show that the 
hyperscaling relation is generally satisfied and the critical index $\eta$ 
generally takes values compatible with 1/4 at the second transition and with 
$4/N^2$ at the first transition, in agreement with the expectations.

\section{Summary} 

This paper completes our study of the critical behavior of $3D$ $Z(N>4)$ 
lattice gauge theories both at finite temperatures.
We have found that in all $Z(N)$ vector models two BKT-like phase 
transitions occur at finite temperatures if $N>4$. In all cases studied, the 
results for the critical indices suggest that finite-temperature $Z(N)$ lattice 
models belong to the universality class of two-dimensional $Z(N)$ vector spin 
models, in agreement with the Svetitsky-Yaffe conjecture. Furthermore, the 
available results for many values of $N$ allowed us to propose and check some
scaling formulas for the critical point of the second phase transition. 
Combining the results of the present paper with those for the index $\nu$ 
obtained by us at zero temperature in Ref.~\cite{ZN_zero_T} enabled us to 
check the continuum scaling and to predict the approximate value for $a T_c$ 
in the continuum limit.

\end{document}